\documentclass[twocolumn]{aastex63}
\usepackage{bm}

\newcommand{\mytilde}{\raise.19ex\hbox{$\scriptstyle\sim$}}

\usepackage[utf8]{inputenc}
\usepackage{CJK}
\usepackage{amsmath}
\usepackage{lineno}
\usepackage{graphicx}
\usepackage{tabularx}

\received{}
\revised{}
\accepted{}

\submitjournal{}

\shortauthors{Cha et al.}

\begin{document}

\title{Weak-lensing Mass Reconstruction of Galaxy Clusters with a Convolutional Neural Network - II: Application to Next-Generation Wide-Field Surveys}

\correspondingauthor{M. James Jee}
\email{sang6199@yonsei.ac.kr, mkjee@yonsei.ac.kr}

\author[0000-0001-7148-6915]{Sangjun Cha}
\affiliation{Department of Astronomy, Yonsei University, 50 Yonsei-ro, Seoul 03722, Korea}
\author[0000-0002-5751-3697]{M. James Jee}
\affiliation{Department of Astronomy, Yonsei University, 50 Yonsei-ro, Seoul 03722, Korea}
\affiliation{Department of Physics and Astronomy, University of California, Davis, One Shields Avenue, Davis, CA 95616, USA}
\author[0000-0003-4923-8485]{Sungwook E. Hong}
\affiliation{Korea Astronomy and Space Science Institute, 776 Daedeok-daero, Yuseong-gu, Daejeon 34055, Korea}
\affiliation{Astronomy Campus, University of Science and Technology, 776 Daedeok-daero, Yuseong-gu, Daejeon 34055, Korea}
\author[0000-0002-6148-2459]{Sangnam Park}
\affiliation{Physics Department \& Natural Science Research Institute, University of Seoul, 163 Seoulsiripdaero, Dongdaemun-gu, Seoul 02504, Korea}
\author[0000-0001-6764-3236]{Dongsu Bak}
\affiliation{Physics Department \& Natural Science Research Institute, University of Seoul, 163 Seoulsiripdaero, Dongdaemun-gu, Seoul 02504, Korea}
\author[0000-0002-6571-4632]{Taehwan Kim}
\affiliation{Artificial Intelligence Graduate School, UNIST, Ulsan, Republic of Korea}

\begin{abstract}
Traditional weak-lensing mass reconstruction techniques suffer from various artifacts, including noise amplification and the mass-sheet degeneracy. In Hong et al. (2021), we demonstrated that many of these pitfalls of traditional mass reconstruction can be mitigated using a deep learning approach based on a convolutional neural network (CNN). In this paper, we present our improvements and report on the detailed performance of our CNN algorithm applied to next-generation wide-field observations. Assuming the field of view ($3\fdg 5 \times 3\fdg5$) and depth (27 mag at $5\sigma$) of the Vera C. Rubin Observatory, we generated training datasets of mock shear catalogs with a source density of 33 arcmin$^{-2}$ from cosmological simulation ray-tracing data. We find that the current CNN method provides high-fidelity reconstructions consistent with the true convergence field, restoring both small and large-scale structures. In addition, the cluster detection utilizing our CNN reconstruction achieves $\sim75$\% completeness down to $\sim 10^{14}M_{\sun}$. We anticipate that this CNN-based mass reconstruction will be a powerful tool in the Rubin era, enabling fast and robust wide-field mass reconstructions on a routine basis.
\end{abstract}


\section{Introduction} 
\label{sec:intro}
Weak lensing (WL) is a powerful method for reconstructing the mass distributions of galaxy clusters without any dynamical assumptions \citep[e.g.,][]{1993ApJ...404..441K,1998MNRAS.299..895B, 1998A&A...337..325S,2000ApJ...532...88H, 2006ApJ...642..720J, 2007ApJ...663..717M, 2011ApJ...729..127U, 2017ApJ...851...46F, 2021ApJ...923..101K, 2024ApJ...973...79A}. 
Despite this unique merit, WL mass reconstruction faces several critical limitations, including the mass-sheet degeneracy, boundary effect, noise amplification, resolution loss, and finite-field effect \citep[e.g.,][]{1985ApJ...289L...1F, Bartelmann:1995yq, Seitz:1995dq, 2004A&A...424...13B}. 

The mass-sheet degeneracy arises because the shear field is invariant under a certain transformation of the mass density. 
The finite-field/boundary effect occurs because the shear field near the edges of the observed region is influenced by the mass outside the reconstruction field, for which we have no constraining data. 
The noise amplification is caused because the mass reconstruction is an ill-posed inversion problem where the relation between the observed shear and underlying mass is nonlinear.
The resolution loss is inevitable because the number density of the sources is finite and thus the shear field needs to be smoothed. 
In general, these issues happen together, and their combined effects are present in mass reconstructions, leading to numerous artifacts and difficulties in interpretation.

One of the notable efforts to mitigate some of the aforementioned artifacts is mass reconstruction through the maximum likelihood approach. In this approach, a grid representing either mass or potential is set up on the reconstruction field, and the 
parameters defining the grid are iteratively determined by maximizing the likelihood relating mass and shear \citep{1996ApJ...464L.115B, 1996ApJ...473...65S, 1999MNRAS.302..118G, 2008ApJ...684..794K}.
However, maximizing the likelihood alone leads to overfitting, and thus it is necessary to impose a prior to regularize the result. Some methods utilizing the entropy of the mass grid as a regularization channel have been demonstrated to be effective in overcoming some pitfalls of traditional mass reconstructions \citep[e.g.,][]{1998MNRAS.299..895B, 1998A&A...337..325S, 2005A&A...437...39B, 2022ApJ...931..127C, 2024ApJ...961..186C}.
One of the major drawbacks of this approach is its large computational cost. 
Even a moderately sized grid of 100×100 requires 10,000 free parameters, making mass reconstruction computationally expensive and slow when using a traditional CPU-based approach.
Hence, it is difficult to apply the method for routine use to deep wide-field (WF) WL data from next-generation facilities such as the Vera C. Rubin Observatory, the Euclid Mission, and the Nancy Grace Roman telescope.

Deep learning is poised to revolutionize many fields of astronomy. In particular, its unparalleled ability to solve ill-posed nonlinear inversion problems holds tremendous potential in mitigating the limitations of traditional approaches in image denoising,  restoration, deconvolution, super-resolution, and more \citep[e.g.,][]{2019PhRvD.100d3527S, 2021MNRAS.504.5543L, 2022MNRAS.517.4054S, 2024ApJ...972...45P}. Certainly, WL mass reconstruction is a promising field that stands to gain from the rapid advancements in deep learning.

Recently, we proposed a convolutional neural network (CNN) architecture for WL mass reconstruction \citep[][hereafter H21]{2021ApJ...923..266H}. The CNN-based mass reconstruction was shown to significantly outperform traditional mass reconstructions in restoring mass density distributions, recovering projected halo masses, suppressing spurious substructures, and predicting mass centroids. 

In this paper, we present our improved CNN architecture and report on the detailed performance when the method is applied to next-generation WF observations. The improvement stems from two factors. First, we enhanced our CNN algorithm to be more compact and efficient, enabling it to preserve the resolution of mass maps for capturing substructures in WF observations. Second, we updated the loss function to ensure that the resulting mass reconstruction captures not only high-contrast individual clusters but also diffuse large-scale structures within the reconstruction field. Among several next-generation WL surveys, we target Vera C. Rubin's Legacy Survey of Space and Time (LSST) and generate training data for its field of view (FOV) ($3\fdg5 \times 3\fdg5$) with the 10-year nominal depth of 27 mag \citep{ivezic2019}. 

The paper is organized as follows. In \textsection\ref{sec:method}, we present our CNN architecture and the methods for generating mock WL data. \textsection\ref{sec:result} shows the reconstruction results. In \textsection\ref{sec:discussion}, we discuss the fidelity of the results with a few metrics. We summarize in \textsection\ref{sec:summary}. Unless stated otherwise, we assume a flat $\Lambda$CDM cosmology with the dimensionless Hubble parameter $h=0.7$ and the matter density parameter $\Omega_{M}=0.3$ throughout the paper.


\section{Method} \label{sec:method}
\subsection{Basic Weak Lensing Theory}\label{WL_for}
In this section, we provide a brief overview of the WL theory. For more details on gravitational lensing, readers can refer to review papers \citep[e.g.,][]{bartelmann2001, hoekstra2013}. 

The distortion of source galaxies in the WL regime can be expressed as $\mathbf{x'}=\mathbf{Ax}$, where $\mathbf{x}$ ($\mathbf{x'}$) is the position in the source (image) plane. The matrix $\mathbf{A}$ is expressed as:
\begin{equation}
    \mathbf{A}= (1-\kappa)
    \begin{pmatrix}
    1-g_1   &   -g_2 \\
    -g_2    &   1 + g_1
    \end{pmatrix}.
    \label{eqn_A}
\end{equation}
In Equation~\ref{eqn_A}, $\kappa$ denotes the convergence, and $g_{1(2)}$ is the first (second) component of the reduced shear $g=({g_1}^2+{g_2}^2)^{1/2}$. The reduced shear $g$ is computed from the relation $g=\gamma/(1-\kappa)$, where $\gamma$ is the shear. The convergence $\kappa$ is defined as:
\begin{equation}
    \kappa=\frac{\Sigma}{\Sigma_c},
\end{equation}
where $\Sigma$ is the surface mass density, and the critical surface mass density $\Sigma_c$ is given by:
\begin{equation}
    \Sigma_c = \frac{c^2 D_s}{4 \pi G D_d D_{ds}}.
\end{equation}
Here, $c$ is the speed of light, $D_{d(s)}$ is the angular diameter distance to the lens (source), and $D_{ds}$ is the angular diameter distance between the lens and the source.
$\gamma$ and $\kappa$ are related by the following convolution:
\begin{equation}
    \bm{\gamma}\mathbf{(x)}=\frac{1}{\pi}\int\mathbf{D}(\mathbf{x}-\mathbf{x'})\kappa(\mathbf{x'})d\mathbf{x'},
\end{equation}
where the convolution kernel $\mathbf{D}$ is:
\begin{equation}
    \mathbf{D}\mathbf{(x)}=-\frac{1}{(x_1 - \mathbf{i}x_2)^2}.
\end{equation}

\subsection{Generation of the Training Data}\label{create_data}
We follow the procedure outlined in H21 for generating the training data. We use the publicly available convergence map from the MassiveNuS cosmological simulation \citep{2018JCAP...03..049L}\footnote{\url{http://columbialensing.org}}. From the convergence map datasets at five different source redshifts ($z=0.5, 1.0, 1.5, 2.0, ~\rm and ~ 2.5$), we select the dataset created for the source plane at $z=2.5$. Note that this $z=2.5$ source redshift of the convergence map is different from the source redshift that we assume for creating the WL galaxy shape catalog described below.
In H21, we selected the convergence map for the source plane at z=1.5; however, we find that using the dataset for the source plane at $z=2.5$ improves the reconstruction quality. Due to the higher lensing efficiency, the convergence map at $z=2.5$ contains richer structures. One potential concern is the risk of bias, as the source plane at $z=2.5$ is significantly higher than the typical value ($z\sim1$) achieved with ground-based data. However, our CNN model is designed to learn the relationship between the shear catalog and the corresponding underlying mass distribution, rather than the characteristic convergence pattern at a specific source redshift. Consequently, we find that this particular choice of source redshift for our training dataset does not introduce a significant bias in the mass reconstruction.
The dataset consists of 10,000 convergence maps covering a field size of $3\fdg5\times3\fdg5$ with a resolution of $512\times512$ ($\sim0\farcm410$ per grid cell).

When the nominal 10-year mission is completed, the LSST is expected to cover the entire southern sky with a depth of 27 mag. From the Subaru Hyper-Suprime Cam (HSC) archival images with the expected median seeing ($\mytilde0.7\arcsec$) of the LSST, we find that the average source number density is $\mytilde$33 ${\rm arcmin}^{-2}$. Since the Subaru/HSC imaging data share many properties with the Rubin Observatory, we adopt this number density as the average source density for our mock data  \citep{2024NatAs.tmp....9H}.
The noise in the shape of each source galaxy is comprised of the shape dispersion and measurement components. As in H21, we adopt $\sigma_e=0.24$ as the intrinsic shape dispersion. For the measurement error, we utilize the relation between galaxy magnitude and measurement error determined from an analysis of Subaru/HSC data in \citet{2024NatAs.tmp....9H}.

We divide the 10000 convergence maps into 7000 for training, 2800 for validation, and 200 for testing. Data augmentation is employed to increase the number of training data by applying four rotations and two-axis flips, but we do not use random crops as done in H21. 
The resulting number of convergence maps for training is $7000\times(4+2)=42000$. Within the 3\fdg5 x 3\fdg5 field, there are no preferred locations for galaxy clusters unlike in H21. 

\subsection{CNN Model Architecture}\label{CNN_model}
\begin{deluxetable}{cccccc}
\tablecaption{Architecture of our convolutional neural network.
\label{tab:architecture}
}
\tablehead {
\colhead{Order} &
\colhead{Layer} &
\colhead{Filter Size} &
\colhead{} &
\colhead{Output Size} 
}
\startdata
1 & Input & - & - & (3, 512, 512) \\
2 & Conv2D-0 & (9, 9) & - & (8, 512, 512) \\
 & Conv2D-1 & (25, 25) & - & (8, 512, 512) \\
 & Conv2D-2 & (49, 49) & - & (8, 512, 512) \\
3 & Concatenate & - & Conv2D-0 & (24, 512, 512) \\
 & &  & Conv2D-1 &  \\
 & &  & Conv2D-2 &  \\
4 & Dropout-0$^1$ & - & - & (24, 512, 512) \\
5 & Conv2D-3 & (49, 49) & - & (24, 464, 464) \\
6 & TransConv2D-0 & (49, 49) & - & (24, 512, 512) \\
7 & Multiply & - & Concatenate & (24, 512, 512) \\
8 & Dropout-1$^1$ & - & - & (24, 512, 512) \\
9 & Conv2D-4 & (49, 49) & - & (24, 464, 464) \\
10 & TransConv2D-1 & (49, 49) & - & (24, 512, 512) \\
11 & Conv2D-5 & (1, 1) & - & (1, 512, 512) \\
12 & Output & - & - & (1, 512, 512)
\enddata
\tablecomments{$^1$Dropout rate is 25 percent. $^2$The number of free parameters in our CNN model is $5.6\times10^6$.}
\end{deluxetable}

\begin{figure*}
\centering
\includegraphics[width=\textwidth]{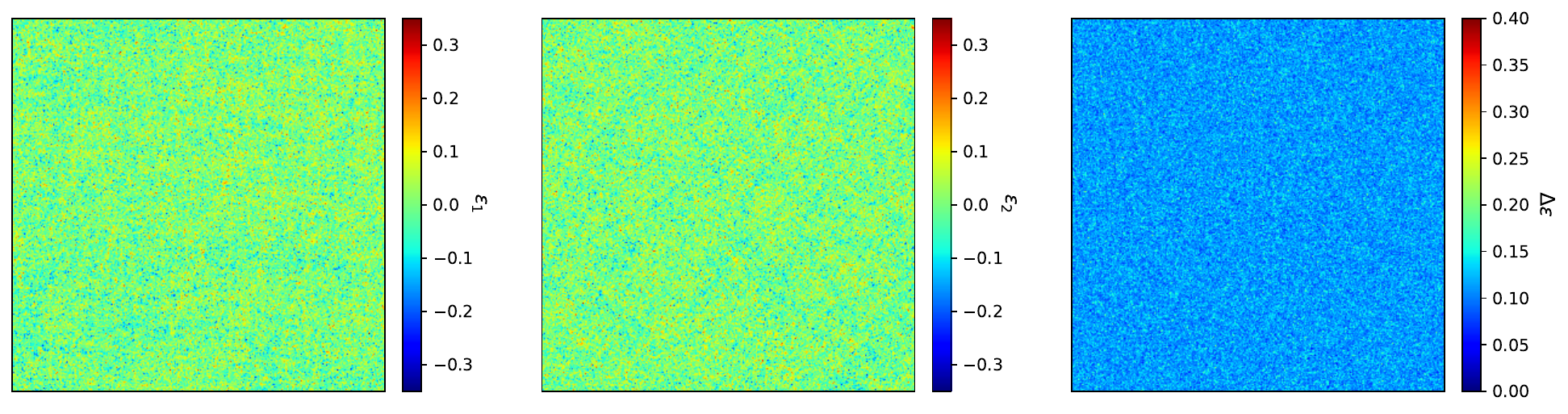} 
\caption{Example of three-channel input layers. The left and middle panels show the $\epsilon_1$ and $\epsilon_2$ channels, respectively, while the right panel displays the $\Delta_\epsilon$ channel.
}
\label{input_maps}
\end{figure*}

Table~\ref{tab:architecture} presents the architecture of our new CNN model. The input layer is the $3\times512\times512$ array consisting of the three ($\epsilon_{1}, \epsilon_{2}$, and $\Delta\epsilon$) channels. The first two channels are the average ellipticity components within the grid cell, while the last channel is the average ellipticity measurement error. Figure~\ref{input_maps} presents an example of the three-channel input layers. 
Unlike in H21, we did not smooth the input layers with a Gaussian kernel\footnote{H21
needed to smooth the input layers because the source density per pixel was much lower than in the current setup.
}. To mitigate overfitting associated with empty pixels, we applied smoothing to the input layers. However, in this study, we did not smooth the input layers because most pixels contain sufficient ellipticity information. 

Our CNN model is composed of 8 two-dimensional (2D) convolutional layers, whereas the previous model used 11 2D convolutional layers. Instead of increasing the depth of our CNN model, we improved it by employing different filter sizes in the initial layers and concatenating them. This approach allows our CNN model to capture both small and large-scale structures with shallower layers than H21, making it more compact and efficient. Additionally, unlike in H21, we maintain the input layer size ($512\times512$) throughout the CNN operations to minimize resolution loss. The number of free parameters in our CNN model is $5.6\times10^6$.

Similarly to the approach in H21, a skip connection is added after each combination of the convolution and transposed convolution layers. This measure is motivated by the residual neural network \citep[ResNet;][]{7780459}, which is shown to improve the overall performance of the CNN model by reducing the problem of vanishing gradients. 
The activation function for the convolutional layer is the Leaky Rectified Linear Activation (LReLU) function, which has a slope of 0.2 when $x<0$.
To prevent overfitting, we applied Dropout layers \citep{JMLR:v15:srivastava14a} after the {\tt Concatenate} and {\tt Multiply} layers, randomly excluding neurons during training. In this study, we excluded 25 percent of neurons per convolution layer. For training our CNN model, the Adam optimizer \citep{adam2014} was used with a learning rate of $10^{-5}$. We ran 1000 epochs and adopted gradient clipping for the optimizer \citep{Zhang2020Why} to prevent exploding gradients.

\subsection{Loss Function}\label{loss_func}
In H21, we utilized a weighted mean-squared-error loss function inspired by the focal loss \citep{Lin_2017_ICCV}. This approach emphasized the importance of higher $\kappa$ values. The loss function in H21 is given by:
\begin{equation}
    L=\sum_{\bm{x}}\omega_{f}(\bm{x})[\kappa_{pred}(\bm{x})-\kappa_{truth}(\bm{x})]^2,
    \label{prev_loss_function}
\end{equation}
where $\omega_{f}(\bm{x})$ is a weight function defined as:
\begin{equation}
    \omega_{f}(\bm{x}) = 1 + \frac{|\kappa_{truth}|}{{\rm max}(\kappa_{truth})}.
    \label{weighted}
\end{equation}

In this study, we aim to reconstruct both high-contrast individual clusters and more diffuse large-scale structures. The ratio of high ($\gtrsim$0.2) to low ($\lesssim$0.1) $\kappa$ pixels in our WF dataset is lower than that in H21, rendering the original weighting scheme less effective for our purposes. Thus, we have modified the loss function to better suit the broader range of structures as follows:
\begin{equation}
    L=\sum_{\bm{x}}{\omega_{f}(\bm{x})}^{6}[\kappa_{pred}(\bm{x})-\kappa_{truth}(\bm{x})]^2,
    \label{new_loss_function}
\end{equation}
where $\omega_{f}(\bm{x})$ is the weight defined in Equation~\ref{weighted}. 

This revision amplifies the influence of high-$\kappa$ pixels more than the original weighting scheme. The choice of the exponent 6 is empirically determined to provide optimal balance and training stability, enhancing the model's sensitivity to high-$\kappa$ pixels while preventing overfitting. By increasing the weight of high-$\kappa$ pixels, the model is better tailored to accurately predict these pixels while still maintaining its ability to reconstruct low-$\kappa$ pixels without overfitting.


\section{Result} \label{sec:result}
To evaluate the performance of our CNN model, we performed mass reconstruction using 200 test datasets that were not used either in training or validation. Additionally, for comparison, we generated mass maps employing the H21 CNN model and the KS93 method \citep{1993ApJ...404..441K}. For the KS93 analysis, we use a smoothing scale of $\sigma\sim150\arcsec$, which is a common choice for smoothing shear data \citep[e.g.,][]{2024arXiv240500115H}. 
When displaying the results from the H21 method, we applied bicubic interpolation to match the dimension of the current results, as the H21 implementation provides lower-resolution outputs. 

\subsection{Visual Analysis}\label{sec:visual_result}
\begin{figure*}
\centering
\includegraphics[width=\textwidth]{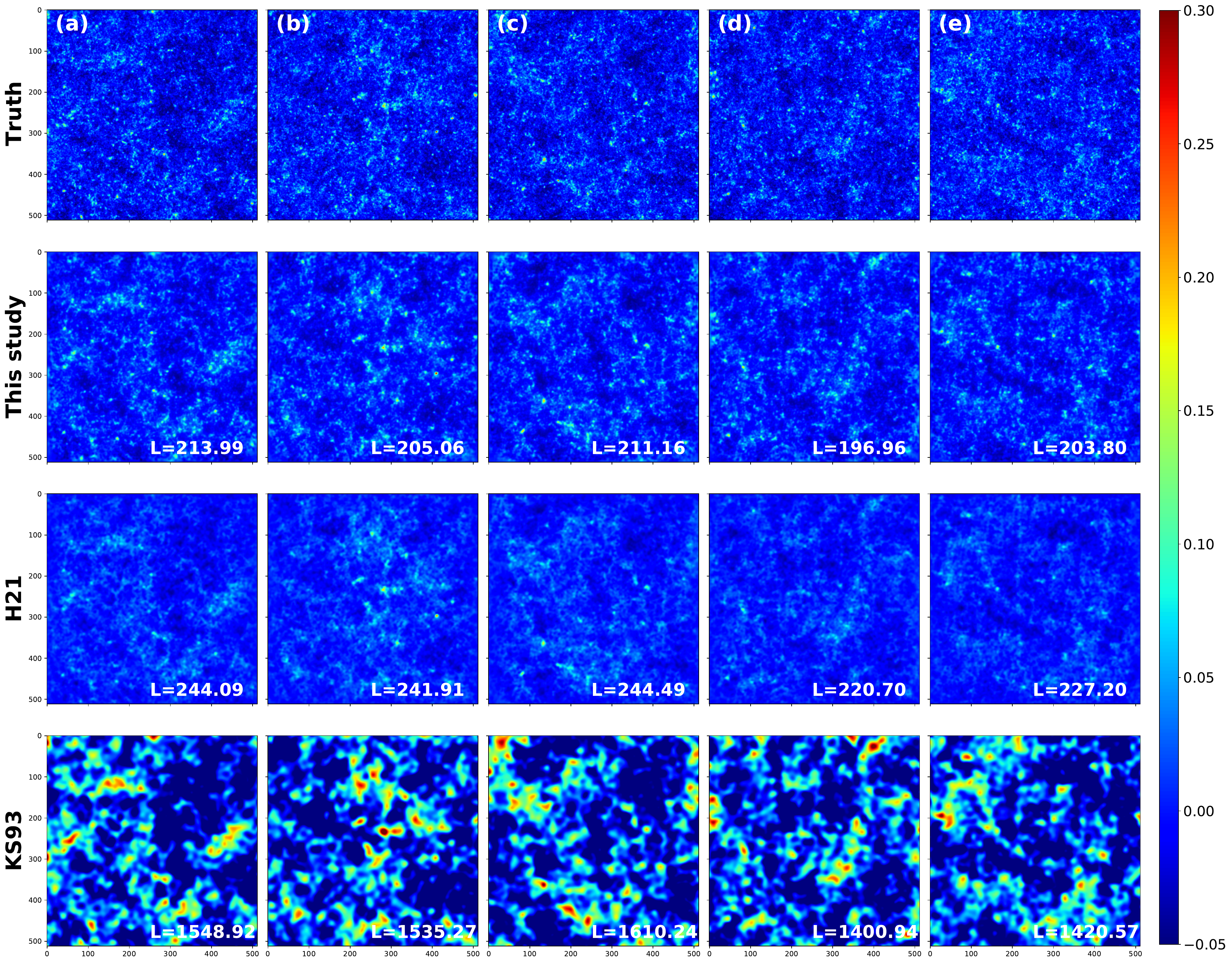} 
\caption{Comparison of the $\kappa$ maps derived from the truth, our CNN model, the KS93 method, and the H21 across five test datasets. The top row displays the truth convergence maps, the second row shows the maps reconstructed by our CNN model, the third row presents the maps reconstructed using the model in H21, and the bottom row exhibits the maps from the KS93 method. $L$ indicates the loss values derived from Equation~\ref{new_loss_function}. Each panel displays a field of $3\fdg5\times3\fdg5$.}
\label{quantitative_result}
\end{figure*}

Figure~\ref{quantitative_result} compares the mass reconstruction results from the current CNN, H21, and KS93 methods for five randomly selected fields from the test dataset. 
The $\kappa$ values from our CNN model range from $\mytilde-0.05$ to $\mytilde0.3$, consistent with the true convergence maps. Notably, the reconstructed maps from our CNN model exhibit good agreement in features with the true $\kappa$ values for both low ($\lesssim$0.1) and high ($\gtrsim$0.2) $\kappa$ values, marking an improvement over the results from the H21 model. In contrast, the KS93 reconstructions tend to significantly overestimate the $\kappa$ values mainly due to the mass-sheet degeneracy.
We also found that the residuals between the truth and reconstructed mass maps from our CNN model do not present any significant systematic pattern.

The model from H21 tends to smooth out small-scale structures, a trend discernible across all samples in Figure~\ref{quantitative_result}. We attribute this to the reduction of the input channels in the first convolutional layer. 
While this approach was introduced to enhance training efficiency, the reduction makes it difficult to capture small-scale structures. 
The KS93 results show correlations with the large-scale structures of the true mass distribution but significantly smooth out finer details.
Of course, one can consider decreasing the kernel size for KS93.
However, when we decreased the kernel size to restore the compactness of the highest peaks, the result showed numerous spurious fluctuations across the field. This is not surprising because the reduced kernel size is inadequate in low density regions.

Finally, the superiority of our new CNN model can also be evaluated by the loss $L$ (see the value in the lower right corner of each panel in Figure~\ref{quantitative_result}) evaluated with Equation~\ref{new_loss_function}.
The $L$ value from our new CNN results spans the range $200-210$, which is $10-20\%$  lower than those from the H21 results. The KS93 result shows significantly higher $L$ values.

\subsection{Direct Comparison of Convergence Values}\label{kappa_value_comp}
\begin{figure}
\epsscale{1.1}
    \plotone{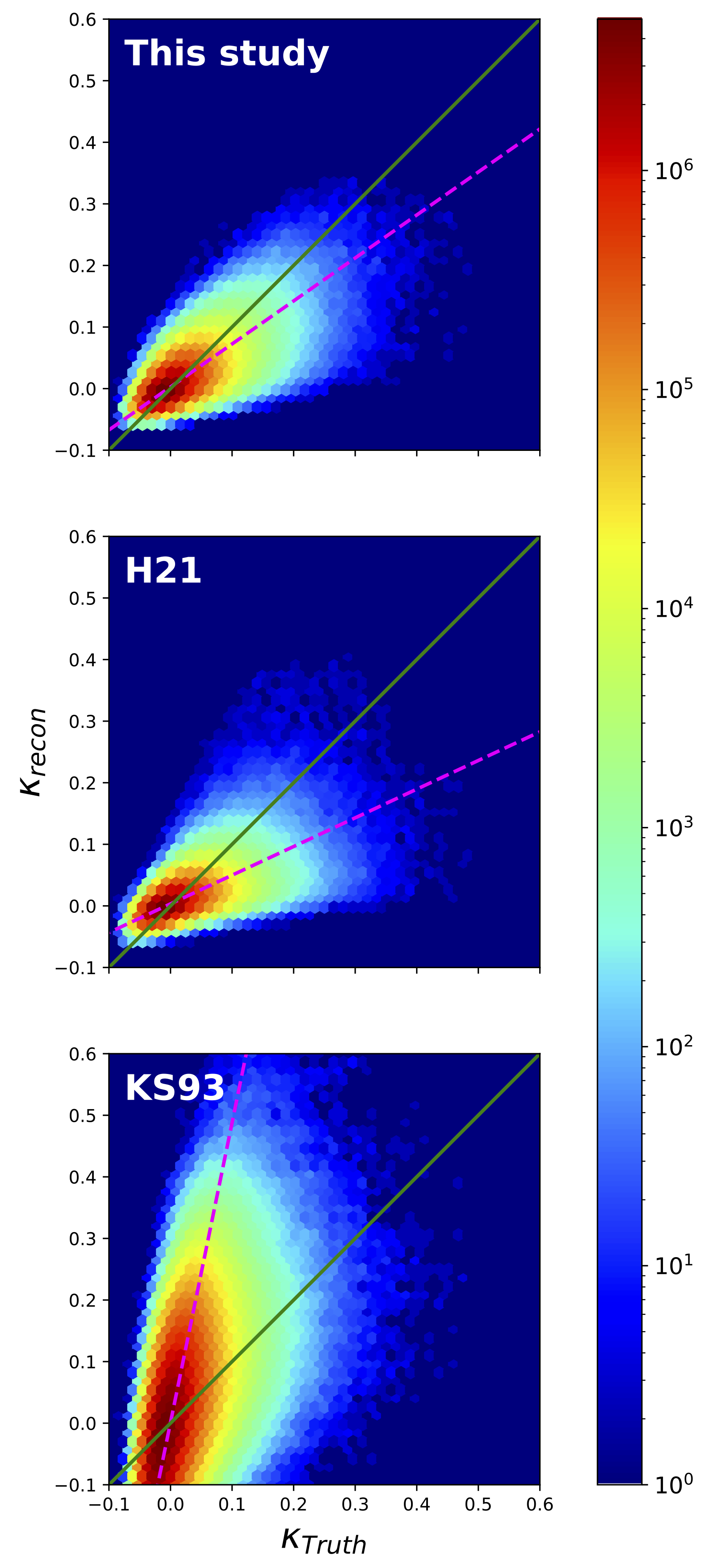}
    \caption{Correlation between the truth and the reconstructed $\kappa$ values pixel-by-pixel. The upper panel (middle and lower panel) shows the correlation for our CNN model (model in H21 and KS93 method). The x-axis and y-axis indicate the truth and the reconstructed $\kappa$ values, respectively. The solid green (dashed magenta) lines represent perfect correlations (total least squares regression results). } 
    \label{one_to_one_correl}
\end{figure}

\begin{figure}
\epsscale{1.1}
    \plotone{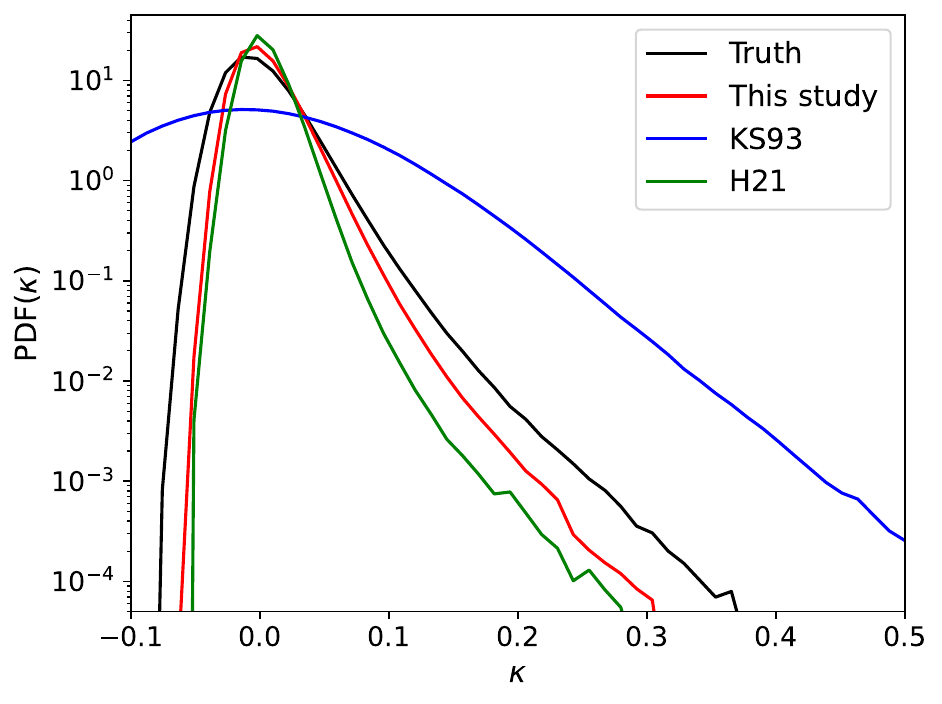}
    \caption{PDF of $\kappa$. The black solid line indicates the PDF of $\kappa$ of the truth maps. The red (blue and green) solid line represents the {probability distribution of} $\kappa$ from our CNN model (KS93 method and model in H21).} 
    \label{kappa_distribution}
\end{figure}
        
Following the qualitative visual analysis in \textsection\ref{sec:visual_result}, we present a quantitative evaluation by directly comparing convergence values.
Figure~\ref{one_to_one_correl} displays pixel-by-pixel correlations between the truth and reconstructed $\kappa$ maps. 
We find that the current CNN results exhibit a stronger correlation, with slopes of $\mytilde0.47$ and $\mytilde0.70$ for the H21 and current models, respectively.
In contrast, the slope deviates substantially from unity for the KS93 mass reconstruction.

Figure~\ref{kappa_distribution} compares the probability distribution function (PDF) of $\kappa$ values.
Once again, it is evident that the distribution derived from the current CNN model matches the truth more closely than the H21 model. In particular, the agreement is significantly improved in the recovery of high $\kappa$ values ($\kappa>0.1$). The distribution is considerably wider in the KS93 reconstruction.

\subsection{$\kappa - \kappa$ Auto-Correlation Functions}\label{kk_corr_func}

\begin{figure}
    \epsscale{1.15}
    \plotone{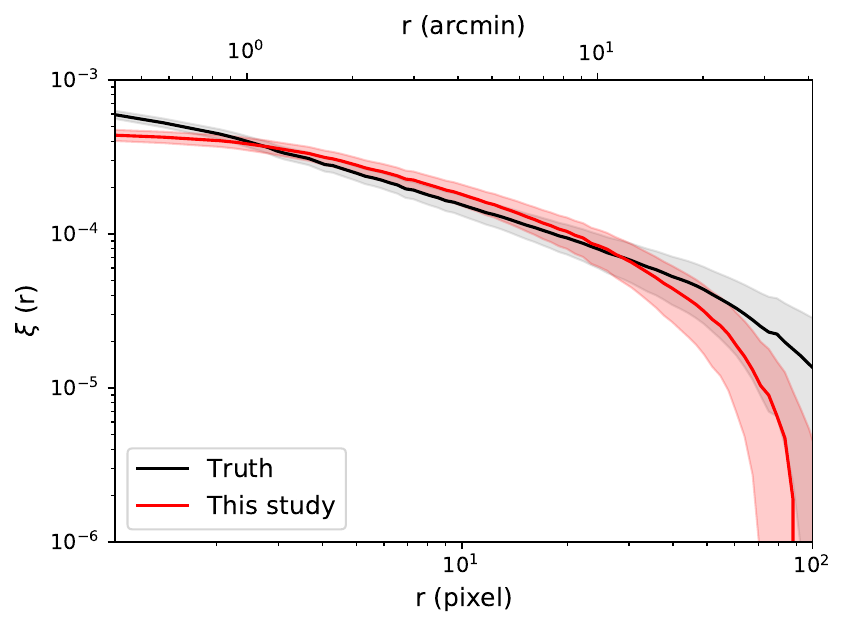}
    \caption{$\kappa-\kappa$ auto-correlation functions. The black (red) solid line shows the median of $\kappa-\kappa$ correlation functions from the 200 truth (reconstructed) convergence maps. The shaded regions indicate 1$\sigma$ uncertainties. 
    The power is well recovered at both small and large scales. The discrepancy at $\lesssim 2$ pixels ($\lesssim0.8$ arcmin) is due to the inevitable resolution loss in the reconstruction caused by the finite source density.
    } 
    \label{k-k_corr}
\end{figure}

In \textsection\ref{sec:visual_result} and \textsection\ref{kappa_value_comp}, we evaluated the fidelity of our CNN mass reconstruction by comparing 2D mass maps, pixel-by-pixel values, and $\kappa$ distributions.
Here, we assess the similarity in structures between the reconstructed map and truth by comparing their $\kappa-\kappa$ auto-correlation functions.

To calculate the auto-correlation function, we used {\tt TreeCorr} \footnote{\url{https://github.com/rmjarvis/TreeCorr}} \citep{2004MNRAS.352..338J, 2015ascl.soft08007J} on 200 mass maps from the truth and the reconstruction.
Figure~\ref{k-k_corr} displays the results.
We find that the correlation is well recovered at both small
and large scales. The discrepancy at $\lesssim 2$ pixels ($\lesssim 0.8$
arcmin) is due to the inevitable resolution loss in the
reconstruction caused by the finite source density.

\subsection{Cluster Detection}\label{cluster_detection}
\begin{figure}
    \epsscale{1.25}
    \plotone{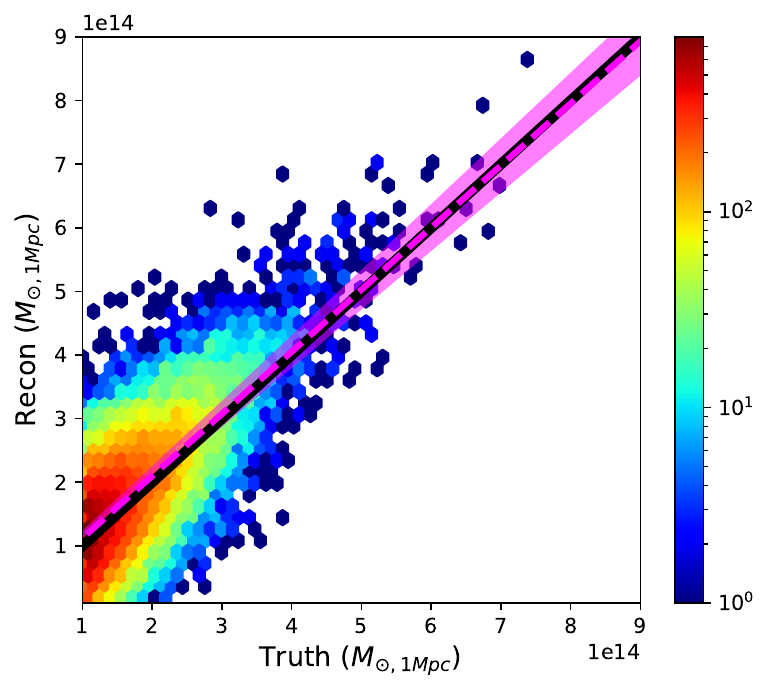}
    \caption{Projected cluster masses within 1 Mpc radius between the truth and reconstruction maps from our CNN model. The black solid (magenta dashed) line indicates a perfect correlation (linear regression) between the truth and reconstructed projected masses. The magenta shade shows the uncertainty.} 
    \label{aperture_mass_comparison}
\end{figure}

\begin{figure}
    \epsscale{1.15}
    \plotone{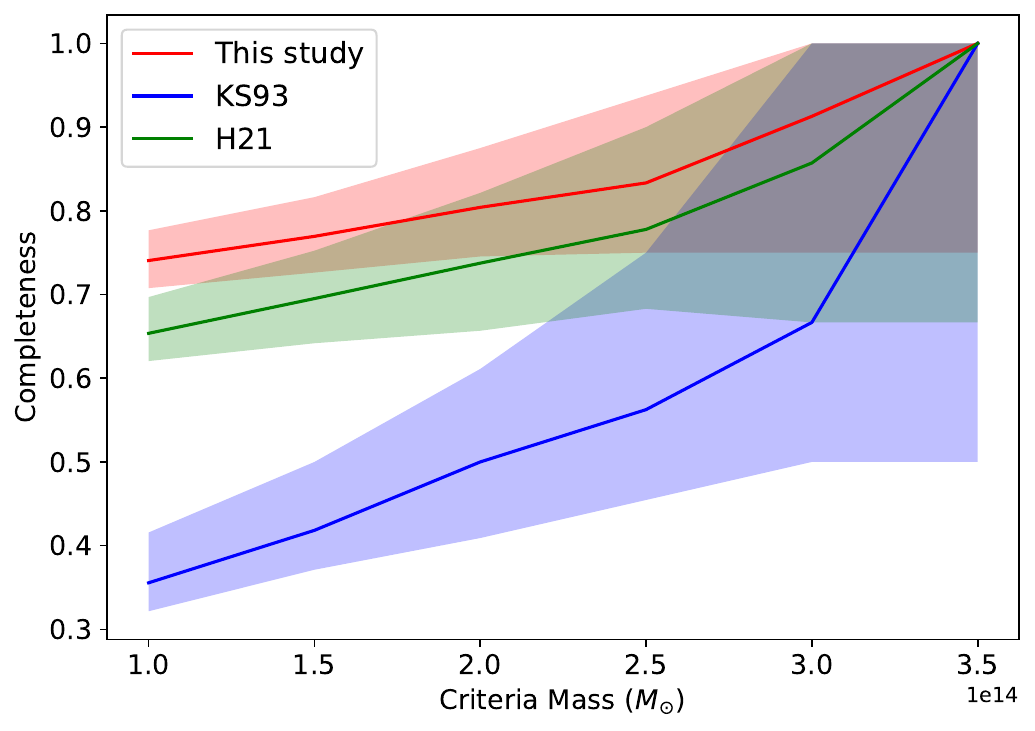}
    \caption{Completeness of the cluster detection. The x-axis and y-axis indicate the cluster mass for detection and completeness, respectively. The red (blue and green) line and shaded region represent the median, and 16th and 84th percentiles of the completeness from our CNN model (KS93 method and model in H21).} 
    \label{completeness}
\end{figure}

One of the key merits of
future WL observations
is their ability to detect galaxy clusters 
solely based on projected mass maps.
This method offers significant advantages over traditional approaches such as galaxy overdensity mapping, Sunyaev-Zeldovich effect observations, and X-ray surveys because the signals directly correlate with their masses.

We assess the cluster detection capability of our CNN model by the following. First, we identify cluster candidates as local peaks on both the truth and reconstructed maps. We then compute projected masses within a 1 Mpc radius for each identified cluster and evaluate the detection performance by comparing the projected mass and completeness. 
Clusters from the truth and CNN results were matched based on a distance criterion, which we set to 6 pixels ($\sim2\farcm4$).

Figure~\ref{aperture_mass_comparison} compares the truth and CNN reconstructed projected cluster masses. 
The slope from the linear regression (dashed magenta line) is very close to the one-to-one correlation (solid black line), indicating that the convergence values obtained directly from the CNN maps serve as accurate mass proxies. 

In Figure~\ref{completeness}, we present the completeness of cluster detection, defined as the ratio of the number of detections in the reconstructed mass maps to those in the true convergence maps, as a function of the cluster mass. All three methods achieve $100 \%$ completeness for clusters with masses larger than $\mytilde 3.5\times10^{14} M_{\odot}$. Our CNN model maintains a  completeness of $\mytilde 75 \%$ down to  $\mytilde 10^{14} M_{\odot}$. The KS93 completeness is much lower, yielding $\mytilde35\%$ at $\mytilde 10^{14} M_{\odot}$. The completeness measured from the H21 result is $\mytilde10$\% lower than that of the current version in the $\lesssim 2.5\times 10^{14}M_{\odot}$ regime.

In the upcoming era of the Vera C. Rubin Observatory, the Roman Space Telescope, and the Euclid mission, 
WF WL mass reconstructions will be performed routinely.
By leveraging the unique property of WL, which eliminates the need for dynamical 
assumptions (i.e., it is only influenced by the gravitational potential), our CNN model would be an effective and independent tool for detecting galaxy cluster candidates. This capability enables a range of studies, such as the cluster mass function reconstruction \citep{2016MNRAS.463.1929M, 2016MNRAS.456.2486D, 2017A&A...608A..65B} and cosmology \citep{2006ApJ...646..881W, 2009ApJ...692.1060V, 2021MNRAS.502.3942H}. In addition, the independently obtained galaxy cluster candidate catalogs will synergize with other cluster surveys across various wavelengths \citep{2015ApJS..216...27B,2022A&A...657A..56K, 2024A&A...685A.106B}.


\section{Discussion} \label{sec:discussion}

\subsection{Null Test}\label{test_null}
\begin{figure}
\epsscale{1.15}
    \plotone{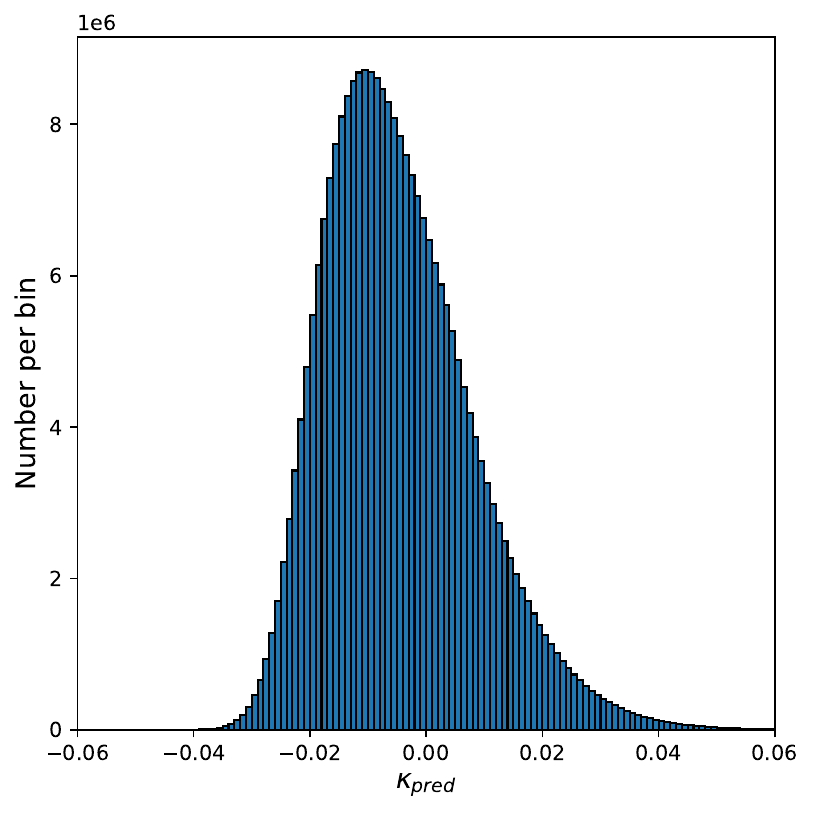}
    \caption{$\kappa$ distribution from the 1000 null fields. The results from the null test have the $\kappa$ values within $-0.02\lesssim\kappa\lesssim0.04$. } 
    \label{null_test}
\end{figure}

In the evaluation of machine learning (ML) models, a null test serves as a critical validation step to ensure that the model does not produce false positive signals from random noise fields.
Since H21 placed clusters always near the center of the field in the training stage, a moderate level of bias was observed for null input data. 

In order to investigate the same issue for the current model,
we generated 1000 mock WL observations from null fields (i.e., $\kappa=0$) with the same observational properties (i.e., including measurement and shape noise) as the training data and reconstructed the corresponding mass maps.
Figure~\ref{null_test} shows the 
$\kappa$ distributions from these mass maps.
The $\kappa$ values 
range between -0.04 and 0.06, and the distribution is positively skewed, with a mode around $-0.01$.
However, compared to H21, the distribution is relatively symmetric, and the bias level is very low.
This improvement is somewhat expected, as the current training dataset is produced from random cosmological ray tracing fields.
Our null test provides confidence in the robustness of the model and its ability to produce reliable mass reconstructions under realistic observational conditions.

\subsection{Boundary \& Finite-Field Effect}\label{boundary_effect}
\begin{figure*}
\centering
\includegraphics[width=\textwidth]{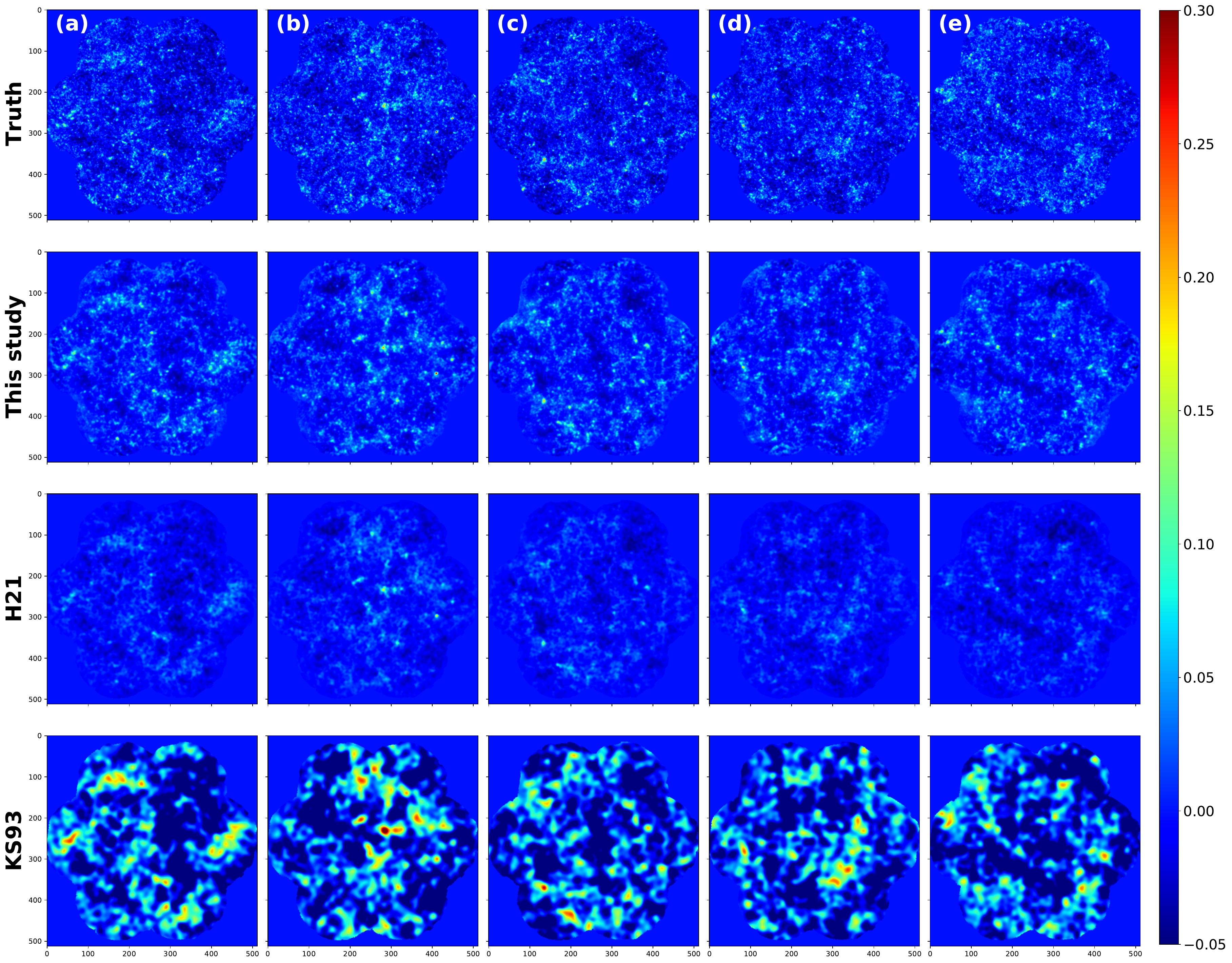} 
\caption{Comparison between the truth and the reconstructed convergence maps using masked WL observations. Same samples are used as Figure~\ref{quantitative_result}.}
\label{masking_result}
\end{figure*}

Traditional WL mass reconstructions are affected by boundary and finite-field effects due to the lack of information beyond the mass reconstruction field. In \textsection\ref{sec:visual_result}, our CNN results do not exhibit these artifacts for square boundaries.

Real observations often have irregular 
observational footprints, which can introduce biases in mass reconstruction.
To investigate these effects under more realistic conditions, we reconstruct $\kappa$ maps with masked input data. 
We adopt the observational footprint of the Subaru/HSC survey of the Coma cluster presented in \citet{2024NatAs.tmp....9H} as our mask.
Figure~\ref{masking_result} shows the resulting mass reconstruction. 

While all three methods can restore the overall mass distributions, we find that the current CNN results are the most satisfactory. Again, the KS93 maps show highly inaccurate convergence values.
The H21 result is significantly better that the KS93 result; However, its convergence values are systematically lower than the true values.
This experiment demonstrates that the current CNN model can be applied to real observations, where complex boundary shapes are present.

\subsection{Application to Real Observations: Mass Reconstruction of the Coma Cluster Field}

\begin{figure*}
\centering
\includegraphics[width=\textwidth]{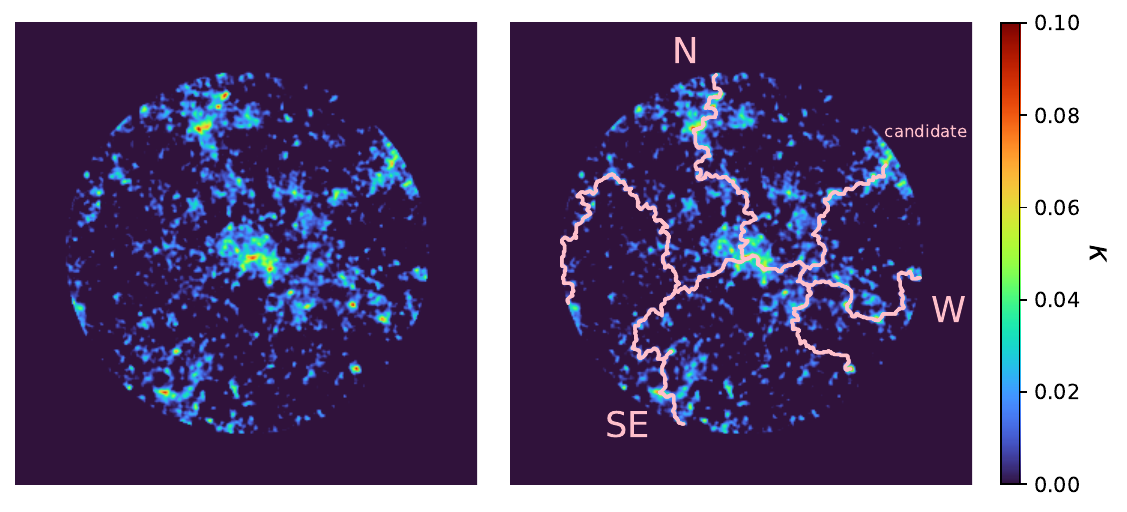} 
\caption{Mass reconstruction of the Coma cluster field. The left panel shows the reconstructed mass map based on the current CNN model, while the right panel overlays the filament candidates (pink solid) detected with {\tt DisPerSE} on the same mass map. The convergence field closely follows the Coma cluster substructures traced by its member galaxies. See text for the discussion on the filament detection.}
\label{coma_mass_map}
\end{figure*}

\begin{figure}
    \epsscale{1.175}
    \plotone{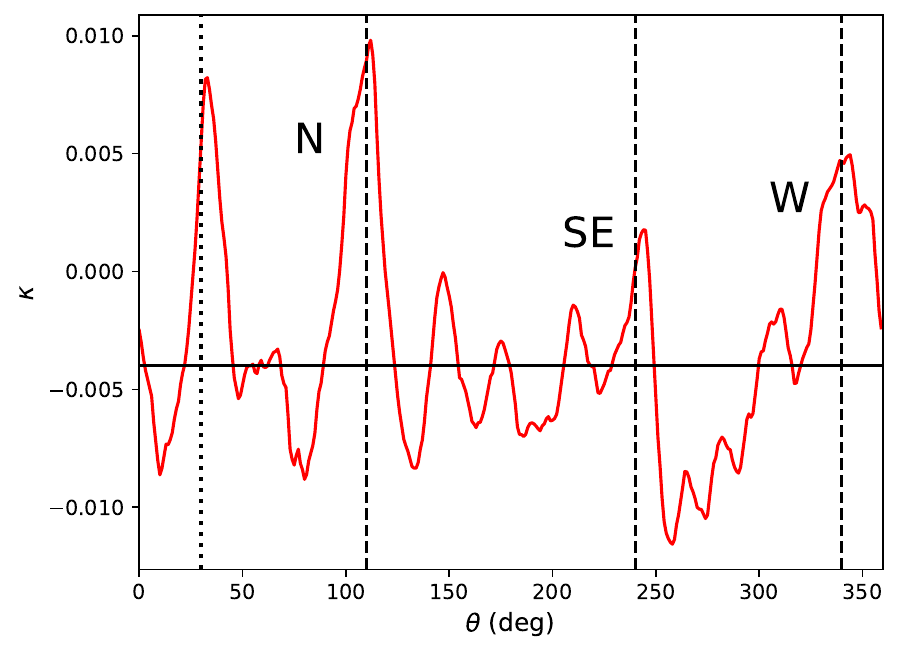}
    \caption{
    Azimuthal distribution of the convergence in the Coma field. 
    The red solid line indicates the mean $\kappa$ within each azimuthal bin.
    The black dashed (dotted) line shows the direction of the intracluster filament (candidate) reported by \cite{2024NatAs.tmp....9H}.  
    }
    \label{azimuthal_kappa}
\end{figure}

While ML has shown remarkable promise in various fields, it is not without its pitfalls. A significant challenge arises when models that perform exceptionally well on training data fail to generalize to real-world applications.
ML models are often sensitive to the quality and representativeness of the training data.  If the training set does not adequately reflect the diversity and complexity of real-world scenarios, the model may become biased or overly simplistic. This lack of robustness can result in poor performance in practical applications, where conditions may vary significantly from those encountered during training. Additionally, changes in the underlying data distribution, known as dataset shift, can further exacerbate these issues, rendering a once-effective model ineffective.

To ensure that the current CNN model excels not only on training datasets but also on real-world data, we experimented with actual WF WL observations. 
We present the mass reconstruction of the Coma cluster in the left panel of Figure~\ref{coma_mass_map}. 
We used the WL shear catalog from \citep{2024NatAs.tmp....9H}, which measured WL signals from Subaru/Hyper-Suprime Cam imaging data covering approximately $3\fdg5\times3\fdg5$ centered on its two BCGs.
The Coma cluster, being a very massive cluster, may serve as an excellent real-world test case, as its characteristics are expected to differ significantly from those of the average cluster in our training dataset.

As the true mass distribution in the Coma field is unknown, we cannot assess its fidelity using the same metrics presented in \textsection\ref{sec:result}. Instead, we provide comparisons with previous results as follows.
First, we find that the mass estimate derived from our convergence field is consistent with previous studies. Fitting an NFW profile provides $M_{200}=7.4_{-3.1}^{2.4}\times10^{14}M_{\sun}$, which aligns with literature values \citep[e.g.,][]{2009A&A...498L..33G, 2014ApJ...784...90O}.
Second, the reconstruction reveals a number of significant substructures, some of which were also reported by previous low-resolution WL studies.
Third, the mass distribution is well traced by those of the Coma cluster galaxies and intracluster light \citep[e.g.,][]{2014ApJ...784...90O, 2024arXiv241215328J}.

Most notably, the CNN mass map reveals some of its intracluster filaments reported by \cite{2024NatAs.tmp....9H}.
Intracluster filaments are the terminal ends of intercluster filaments that stretch over tens of megaparsecs. Due to their low contrast, their detection has been considered elusive. 
\cite{2024NatAs.tmp....9H} identified the Coma cluster's intracluster filaments not through mass reconstruction, but using two novel methods: the matched-filter technique and the shear peak statistic \citep{Maturi2013}.
Here, we independently detect the intracluster filament of the Coma based on the current CNN mass map as follows. First, we ran {\tt DisPerSE}\footnote{\url{https://www2.iap.fr/users/sousbie/web/html/indexd41d.html}} \citep{2011MNRAS.414..350S, 2011MNRAS.414..384S} on the CNN mass map.
To set the threshold cut for filament detection, we adopt a persistence above the 3-sigma level, where persistence indicates the absolute difference between the saddle and maximum critical points. The solid pink lines in the right panel of Figure~\ref{coma_mass_map} depict the resulting filaments.
The three main branches originating from the cluster center is clearly detected and aligns well with the result from \cite{2024NatAs.tmp....9H}.
Second, we measured the azimuthal excess of the convergence with respect to the background value. Figure~\ref{azimuthal_kappa} shows that this method too reveals the three main filaments identified by {\tt DisPerSE}, which is also consistent with the findings of \cite{2024NatAs.tmp....9H} (compare with Figure 2 of their paper).
The consistency in intracluster filament detection showcases the utility of CNN-based WL mass reconstruction for identifying large-scale structures in future WL surveys, including the detection of cosmic webs. We refer readers to \citet{2024NatAs.tmp....9H} for further scientific details on the Coma cluster.


\section{Summary \& Conclusion} \label{sec:summary}
In this paper, we have presented an improved multi-layered CNN model
for WF WL mass reconstruction and evaluated its performance using mock next-generation WF observations. 
Our enhancements include a more compact and efficient CNN architecture and a refined loss function that targets the recovery of both high-contrast individual clusters and diffuse large-scale structures.

We find that the current CNN model significantly outperforms the H21 version across several metrics.
Visual comparisons of the 2D convergence maps demonstrate that both large-scale
structures and high-density cluster peaks have been remarkably well restored.
Pixel-by-pixel comparisons reveal a stronger correlation, and
the convergence distribution obtained from the current CNN model aligns closely with the truth. However, improvements are still needed at both extremes of the distribution.

Additionally, we demonstrate that our current CNN model serves as a reliable tool for detecting galaxy clusters based solely on gravitational lensing data. The detection completeness reaches 75\% at $\mytilde10^{14}M_{\odot}$ and 90\% at $\mytilde 3\times 10^{14}M_{\odot}$.
The projected cluster masses estimated from the mass map show a near one-to-one correlation with the true values.

Our CNN model maintains robust performance even when the field boundaries are irregular. While the convergence scales near the field boundaries are severely distorted in the H21 result, artifacts in our new mass reconstruction are negligible.

Finally, we use Subaru/HSC WF WL data of the Coma cluster to verify the robustness of our CNN model when applied to real-world data.
The Coma cluster, as a very massive system, provides an excellent real-world test case since its characteristics are expected to differ significantly from those of the average cluster in our training dataset.
We find that the mass estimates and distributions are highly consistent with literature values and luminous tracers.
We also demonstrate that the CNN mass map alone can be used to identify  intracluster filaments reported by independent methods such as the matched-filter statistic and shear-peak count.

Current and future WL surveys will cover a significant fraction of the sky, necessitating high-fidelity mass reconstruction on a routine basis with arbitrary observational footprints.
We anticipate that our CNN method will be a valuable tool, enabling fast and robust WF mass reconstructions.

We thank the Columbia Lensing group (\url{http://columbialensing.org}) for their simulations available. The creation of these simulations is supported through grants NSF AST-1210877, NSF AST-140041, and NASA ATP-80NSSC18K1093. We thank New Mexico State University (USA) and Instituto de Astrofisica de Andalucia CSIC (Spain) for hosting the Skies \& Universes site for cosmological simulation products. 
This work was supported by the Institute of Information \& communications Technology Planning \& Evaluation (IITP) grant funded by the Korea government (MSIT; No.2021-0-02068, Artificial Intelligence Innovation Hub). M. J. Jee acknowledges support for the current research from the National Research Foundation (NRF) of Korea under the programs 2022R1A2C1003130 and RS-2023-00219959. SC acknowledges this research was supported by Basic Science Research Program through the National Research Foundation of Korea (NRF) funded by the Ministry of Education (No. RS-2024-00413036). S.E.H. is supported by the Korea Astronomy and Space Science Institute grant funded by the Korea government (MSIT) (No. 2024186901). SP and DB were supported in part by NRF Grant RS-2023-00208011, and by Basic Science Research Program through NRF funded by the Ministry of Education (2018R1A6A1A06024).

\software{Astropy \citep{astropy2013,2018AJ....156..123A,2022ApJ...935..167A}, DisPerSE \citep{2011MNRAS.414..350S, 2011MNRAS.414..384S}, Matplotlib \citep{matplotlib2007}, NumPy \citep{harris2020array}, SExtractor \citep{bertin1996}, SciPy \citep{scipy2020}, Tensorflow \citep{tensorflow2015-whitepaper}, TreeCorr \citep{2004MNRAS.352..338J, 2015ascl.soft08007J}}

\bibliographystyle{apj}
\bibliography{main}

\end{document}